\documentstyle[12pt]{article}
\begin{document}
\newcommand{\dfrac}[2]{\frac{\displaystyle #1}{\displaystyle #2}}
\baselineskip=12pt
\title{\bf
Two-point function reduction of four-point amputated functions and transformations 
in $F\bar{F}$ and $RA$ basis in a real-time finite temperature NJL model\thanks{This 
work is partially supported by the National Natural Foundation of China and by Grant 
No.LWTZ-1298 of the Chinese Academy of Sciences.}
 \\}
\author{
{\bf
Bang-Rong Zhou
} \\
\normalsize Department of Physics, The Graduate School of The Chinese Academy of Sciences \\
\normalsize Beijing 100039, China \\
\normalsize and CCAST (World Laboratory), P.O. Box 8730, Beijing 100080, China \\
}
\date{}
\maketitle 
\begin{abstract}
Based on a general analysis of Green functions in the real-time thermal field theory, 
we have proven that the four-point amputated functions in a NJL model in the fermion 
bubble diagram approximation behave like usual two-point functions. We expound the 
thermal transformations of the matrix propagator for a scalar bound state in the 
$F\bar{F}$ basis and in the $RA$ basis. The resulting physical causal, advanced and 
retarded propagator are respectively identical to corresponding ones derived in the 
imaginary-time formalism and this shows once again complete equivalence of the two 
formalisms of thermal field theory on the discussed problem in the NJL model.
\end{abstract}

\\ 
\section{Introduction}
For discussion of the Nambu-Goldstone mechanism [1] of spontaneous symmetry breaking 
at finite temperature in a Nambu-Jona-Lasinio (NJL) model [2], one must calculate 
the propagators for scalar bound states based on four-point amputated Green functions 
[3,4]. In the fermion bubble diagram approximation, the calculation of a four-point 
amputated function in a NJL model at finite temperature can be effectively reduced 
to the one of a two-point function.  This is obvious in the imaginary-time formalism 
[5] of thermal field theory [4]. However, in the real-time formalism [6,7], although 
the obtained four-point amputated functions show a structure of $2\times 2$ matrix 
and one can simply diagonalize it by a $2\times 2$ thermal transformation matrix, 
the derived thermal matrix is a mere effective one and has not yet been explained 
from fundamental principles of thermal field theory. In this paper, we will examine 
theoretical origin of the above reduction.  Starting from the general transformation 
of thermal matrix propagator in the real-time formalism, we will prove that the 
calculations of the four-point amputated functions in a NJL model in the fermion 
bubble diagram approximation can be indeed reduced to the ones of conventional 
two-point functions in the following meanings: 1) the thermal transformation of the 
four-point amputated functions is identical to the one of usual two-point functions 
and 2) the relations among the thermal components of the four-point amputated 
functions can be changed into the ones among the thermal components of corresponding 
two-point functions. Since the thermal transformations of the real-time Green 
functions can be generally made in different bases e.g. $F\bar{F}$ basis, $RA$ basis 
and Keldysh basis etc. [8,9], the reduction of the four-point amputated functions 
to two-point functions will make the above transformations become very simple and 
feasible. \\
\indent The paper is arranged as follows. In Sect.2 we will prove the reduction stated above of the four-point amputated functions in a NJL model to usual two-point 
functions. In Sect.3 and Sect.4, as an example, we will discuss thermal transformation 
of the matrix propagator for a scalar bound state respectively in the $F\bar{F}$ basis 
and in the $RA$ basis and compare the derived physical propagators with the 
corresponding ones obtained in the imaginary-time formalism. Finally, in Sect.5 we 
come our conclusions. \\
\section{Two-point function reduction of four-point amputated functions in NJL model}
Consider a chiral $U_L(1)\times U_R(1)$ NJL model in the real-time formalism with 
the four-fermion interactions
\[
{\cal L}^R_{4F}=
\frac{G}{4}\sum_{a=1}^{2}\{[{(\bar{\psi}\psi)}^{(a)}]^2-[{(\bar{\psi}\gamma_5\psi)}^{(a)}]^2\}(-1)^{a+1}, 
\]
where $\psi$ represents the fermion field with single flavor and $N$ colors and $G$ is the coupling constant; $a=1$ denotes physical fields and $a=2$ ghost 
fields.  Denote the four external four-momenta corresponding to a four-point 
function respectively by incoming $p_1$ and $p_2$ and outgoing $p_3$ and $p_4$, then 
the four-point functions with four external propagators (legs) may have the following 
three forms:
\[G_4(p_1,p_2,-p_3,-p_4), \ \ G_4(p_2,-p_3,-p_4,p_1) \ {\rm and} \ G_4(p_2,-p_4,-p_3,p_1), \]
where the momentum with the minus sign "$-$" means outgoing, otherwise incoming [9].
In the fermion bubble diagram approximation, the external legs corresponding to the 
left two momenta and the right two momenta in the arguments of $G_4$ will separately 
intersect at a common vertex.  \\
\indent Let us consider $ G_4(p_1,p_2,-p_3,-p_4)$ first.  It can be expressed by
\begin{eqnarray}
G_4^{a_1a_2b_3b_4}(p_1, p_2, -p_3, -p_4)&=&[iS(p_1)]^{a_1a}[iS(p_2)]^{a_2a}
\Gamma_4^{aabb}(p_1, p_2, -p_3, -p_4)\nonumber \\
&&\times[iS^T(-p_3)]^{bb_3}[iS^T(-p_4)]^{bb_4},
\end{eqnarray}
where $[iS(p)]^{a'a}$ is the elements of the thermal matrix propagator for free 
fermion, $[iS^T(-p)]^{b'b}=[iS(-p)]^{bb'}$, i.e. "$T$" denotes transpose of the
matrix, and $\Gamma_4^{aabb}(p_1, p_2, -p_3, -p_4)$ represent the corresponding 
four-point amputated functions.  Bearing in mind the special form of summing the 
indices in Eq. (1), we may compactly write Eq. (1) in the matrix form
\begin{equation}
G_4(p_1, p_2, -p_3, -p_4)=iS(p_1)\otimes iS(p_2)
\Gamma_4(p_1, p_2, -p_3, -p_4)iS^T(-p_3)\otimes iS^T(-p_4),
\end{equation}
where $"\otimes"$ denotes external product of the matrices. Following the method 
in Ref. [9], we can make a general transformation of the matrix propagator $iS(p)$
and obtain the transformed matrix propagator 
\begin{equation}
i\hat{S}(p)=U(p)iS(p)U^T(-p),
\end{equation}
where $U(p)$ is a $2\times 2$ matrix.  A special form of $U(p)$ is the thermal 
transformation matrix in usual real-time formalism which diagonalizes $iS(p)$ and 
leads to the physical causal propagator for free fermion and its complex conjugate.
From Eq. (3) it follows that
\begin{equation}
i{\hat{S}}^T(-p)=U(p)iS^T(-p)U^T(-p).
\end{equation}
Now left multiply $U(p_1)\otimes U(p_2)$ and right multiply $U^T(-p_3)\otimes 
U^T(-p_4)$ in Eq. (2) then by means of Eqs. (3) and (4), we can obtain the transformed
four-point function matrix with external legs
\begin{equation}
\hat{G}_4(p_1, p_2, -p_3, -p_4)=i\hat{S}(p_1)\otimes i\hat{S}(p_2)
\hat{\Gamma}_4(p_1, p_2, -p_3, -p_4)i\hat{S}^T(-p_3)\otimes i\hat{S}^T(-p_4),
\end{equation}
where
\[
\hat{\Gamma}_4(p_1, p_2, -p_3, -p_4)=
V(p_1)\otimes V(p_2) {\Gamma}_4(p_1, p_2, -p_3, -p_4) V^T(-p_3)\otimes V^T(-p_4)
\]
is the transformed four-point amputated functions and the denotations
\[
V(p)=[U^T(-p)]^{-1}, \ \ V^T(-p)=U^{-1}(p)
\]
has been used. When writing Eq. (5) through its components we must note that, as far 
as the discussed four-point functions in a NJL model are concerned, the transformed 
external legs $i\hat{S}(p_1)$ and $i\hat{S}(p_2)$ will intersect at a same vertex 
and $i\hat{S}^T(-p_3)$  and $i\hat{S}^T(-p_4)$ will also intersect at another same 
vertex, thus we can obtain
\begin{eqnarray*}
\hat{G}_4^{a_1a_2b_3b_4}(p_1, p_2, -p_3, -p_4)&=&
[i\hat{S}(p_1)\otimes i\hat{S}(p_2)]^{a_1a_2a'a'}
\hat{\Gamma}_4^{a'a'b'b'}(p_1, p_2, -p_3, -p_4)\nonumber \\
&&\times [i\hat{S}^T(-p_3)\otimes i\hat{S}^T(-p_4)]^{b'b'b_3b_4},
\end{eqnarray*}
where the transformed four-point amputated function will be
\begin{equation}
\hat{\Gamma}_4^{a'a'b'b'}(p_1, p_2, -p_3, -p_4)=
[V(p_1)\otimes V(p_2)]^{a'a'aa} {\Gamma}_4^{aabb}(p_1, p_2, -p_3, -p_4) 
[V^T(-p_3)\otimes V^T(-p_4)]^{bbb'b'}.
\end{equation}
Since the four indices of all the matrices in Eq. (6) can always be divided into two 
pairs each with same entries, Eq. (6) will be reduced to the form of $2\times 2$ matrix. 
Noting that, owing to the four-momentum conservation , $\Gamma_4(p_1, p_2, -p_3, -p_4)$ only depend on $p=p_1+p_2=p_3+p_4$.  Thus we can denote
\begin{eqnarray}
\hat{\Gamma}_4^{a'a'b'b'}(p_1, p_2, -p_3, -p_4)&\equiv &{\hat{\Gamma}}^{a'b'}(p), \nonumber \\
{[V(p_1)\otimes V(p_2)]}^{a'a'aa}&\equiv &{[O(p)]}^{a'a}, \nonumber \\
{[V^T(-p_3)\otimes V^T(-p_4)]}^{bbb'b'}&\equiv &{[O^T(-p)]}^{bb'}, \nonumber \\
{\Gamma}_4^{aabb}(p_1, p_2, -p_3, -p_4)&\equiv &\Gamma^{ab}(p)
\end{eqnarray}
and then obtain from Eq. (6)
\[{\hat{\Gamma}}^{a'b'}(p) =[O(p)]^{a'a}\Gamma^{ab}(p) [O^T(-p)]^{bb'} \]
or in the form of $2\times 2$ matrix 
\begin{equation}
{\hat{\Gamma}}(p)=O(p)\Gamma(p)O^T(-p) \ \ {\rm with} \ \ p=p_1+p_2=p_3+p_4.
\end{equation}
Eq. (8) indicates that the transformation of the four-point amputated function 
corresponding to $ G_4(p_1, p_2, -p_3, -p_4)$ is indeed effective to the one of a 
two-point function $\Gamma(p)$.  In this case, $p^2=(p_1+p_2)^2>0$, i.e. $p$ is a 
time-like four-momentum. \\
\indent We can make similar discussions for $ G_4(p_2,-p_3, -p_4, p_1)$ and $ G_4(p_2, -p_4, -p_3, p_1)$.  For instance, the transformed function of $G_4(p_2, -p_3, -p_4, p_1)$ will be expressed by 
\[
\hat{G}_4(p_2, -p_3, -p_4, p_1)=i\hat{S}(p_2)\otimes i\hat{S}(-p_3)
\hat{\Gamma}_4(p_2, -p_3, -p_4, p_1)i\hat{S}^T(-p_4)\otimes i\hat{S}^T(p_1),
\]
where the transformed four-point amputated function $\hat{\Gamma}_4(p_2, -p_3, -p_4, p_1)$ have the following component form :
\begin{equation}
\hat{\Gamma}_4^{a'a'b'b'}(p_2, -p_3, -p_4, p_1)=
[V(p_2)\otimes V(-p_3)]^{a'a'aa} {\Gamma}_4^{aabb}(p_2, -p_3, -p_4, p_1) 
[V^T(-p_4)\otimes V^T(p_1)]^{bbb'b'}.
\end{equation}
In this case, ${\Gamma}_4(p_2, -p_3, -p_4, p_1)$ and $\hat{\Gamma}_4(p_2, -p_3, -p_4, p_1)$ are only the function of $p=p_2-p_3=p_4-p_1$, thus Eq. (9) can be effectively written by
\begin{equation}
{\hat{\Gamma}}^{a'b'}(p) =[O(p)]^{a'a}\Gamma^{ab}(p) [O^T(-p)]^{bb'} 
\end{equation}
with 
\begin{equation}
O(p)\equiv V(p_2)\otimes V(-p_3), \ \ O^T(p)\equiv V^T(-p_4)\otimes V^T(p_1), 
\end{equation}
however, now $p^2<0$, i.e. $p$ is a space-like four-momentum. For $G_4(p_2, -p_4, -p_3, p_1)$, we have the similar conclusion. In fact, if exchanging $-p_3$ with $-p_4$ in $\hat{G}_4(p_2, -p_3, -p_4, p_1)$ we will obtain the transformed $\hat{G}_4(p_2, -p_4, -p_3, p_1)$, the corresponding four-point amputated function $\hat{\Gamma}(p)$
will have similar effective transformation of $2\times 2$ matrix form to Eq. (10), 
where $p=p_2-p_4=p_3-p_1$ with $p^2<0$, i.e. $p$ is also a space-like momentum.\\
\indent The above discussions show that no matter time-like or space-like the momentum 
$p$ is, the thermal transformation of the four-point amputated functions in a NJL 
model in the fermion bubble diagram approximation can always be effectively reduced 
to the one of two-point functions. \\
\indent We may further prove that the relations among the thermal components of the 
above four-point amputated function will also be changed into the ones among the 
thermal components of the corresponding two-point function. The four-point amputated 
functions $\Gamma^{a_1a_2a_3a_4}(p_1,p_2,p_3,p_4) \ (a_1,a_2,a_3,a_4=1,2)$ 
generally obey the following relations [9-11]:
\begin{equation}
\sum_{a_i=1,2}\Gamma_4^{a_1\cdots a_4}(p_1,\cdots, p_4)\prod_{a_i=2}e^{\sigma p^0_i}=0,
\end{equation}
\begin{equation}
\sum_{a_i=1,2}\Gamma_4^{a_1\cdots a_4}(p_1,\cdots, p_4)
\prod_{a_i=2}\eta_i e^{\sigma p^0_i-x_i}=0,
\end{equation}
where $x=\beta(p^0-\mu)$, $\mu$ is the chemical potential and $\eta=+1(-1)$ for boson 
(fermion). If one $p_i$ in $\Gamma_4^{a_1\cdots a_4}(p_1,\cdots, p_4)$ is written 
$-p_i$ (incoming is changed into outgoing), then the corresponding product factor 
in $\prod_{a_i=2}$ in Eq. (13) should be replaced by $\eta_i e^{x_i-\sigma p^0_i}$.
$\sigma$ is an arbitrary parameter, $0\leq\sigma\leq \beta$ [6,7]. In the following
$\sigma=\beta/2$ will be specified. We will consider in turn the above four-point 
amputated functions ${\Gamma}_4^{aabb}(p_1, p_2, -p_3, -p_4)$, 
${\Gamma}_4^{aabb}(p_2, -p_3, -p_4, p_1)$ and ${\Gamma}_4^{aabb}(p_2, -p_4, -p_3, 
p_1)$.  For ${\Gamma}_4^{aabb}(p_1, p_2, -p_3, -p_4)$, Eq. (12) becomes
\[\Gamma_4^{1111}(p_1, p_2, -p_3, -p_4)
+\Gamma_4^{1122}(p_1, p_2, -p_3, -p_4)e^{-\frac{\beta}{2}(p^0_3+p^0_4)}
+\Gamma_4^{2211}(p_1, p_2, -p_3, -p_4)e^{\frac{\beta}{2}(p^0_1+p^0_2)} \]
\begin{equation}
+\Gamma_4^{2222}(p_1, p_2, -p_3, -p_4)e^{\frac{\beta}{2}(p^0_1+p^0_2-p^0_3-p^0_4)}=0.
\end{equation}

In view of $p=p_1+p_2=p_3+p_4$ and that $\Gamma_4^{aabb}$ is only function of $p$, 
we can obtain from Eq. (14) that
\begin{equation}
\Gamma^{11}(p)+\Gamma^{12}(p)e^{-\beta p^0/2}+\Gamma^{21}(p)e^{\beta p^0/2}+
\Gamma^{22}(p)=0,
\end{equation}
where the definition ${\Gamma}_4^{aabb}(p_1, p_2, -p_3, -p_4)\equiv \Gamma^{ab}(p)
$ in Eq. (7) has been used.  Next, Eq. (13) becomes
\begin{eqnarray}
\Gamma_4^{1111}(p_1, p_2, -p_3, -p_4)&+&\Gamma_4^{1122}(p_1, p_2, -p_3, -p_4)e^{x_3+x_4-\frac{\beta}{2}(p^0_3+p^0_4)}\nonumber \\
&+&\Gamma_4^{2211}(p_1, p_2, -p_3, -p_4)e^{\frac{\beta}{2}(p^0_1+p^0_2)-x_1-x_2}\nonumber \\
&+&\Gamma_4^{2222}(p_1, p_2, -p_3, -p_4)e^{\frac{\beta}{2}(p^0_1+p^0_2-p^0_3-p^0_4)-x_1-x_2+x_3+x_4}=0.
\end{eqnarray}
Let $p_1$, $p_4$ correspond to fermions, $p_2$, $p_3$ to antifermions, then we will 
have $\mu_3=-\mu=-\mu_4$, $\mu_1=\mu=-\mu_2$ and
\[x_3+x_4-\frac{\beta}{2}(p^0_3+p^0_4)=\frac{\beta}{2}p^0,\]
\[
\frac{\beta}{2}(p^0_1+p^0_2)-x_1-x_2=-\frac{\beta}{2}p^0,
\]
hence Eq. (16) is reduced to 
\begin{equation}
\Gamma^{11}(p)+\Gamma^{12}(p)e^{\beta p^0/2}+
\Gamma^{21}(p)e^{-\beta p^0/2}+\Gamma^{22}(p)=0.
\end{equation}
This proves that the relations (12) and (13) for the four-point amputated functions 
${\Gamma}_4^{aabb}(p_1, p_2, -p_3, -p_4)$ are changed into the relations (15) and 
(17) obeyed by the two-point functions $\Gamma^{ab}(p)$ with $p^2>0$. \\
\indent By similar way we can prove that the relations (12) and (13) for the four-point 
amputated functions ${\Gamma}_4^{aabb}(p_2, -p_3, -p_4, p_1)$ and 
${\Gamma}_4^{aabb}(p_2, -p_4, -p_3, p_1)$ will also be reduced to the same relations 
(15) and (17) obeyed by the two-point functions $\Gamma^{ab}(p)$, but in the two cases 
we will have $p^2<0$. Therefore, the relations among the four-point amputated 
functions ${\Gamma}_4^{aabb}$ in a NJL model may always be expressed by the relations 
(15) and (17) among the two-point functions $\Gamma^{ab}(p)$, no matter whether the 
squared transfer momentum $p^2>0$ or $p^2<0$. It is indicated that if 
$\Gamma^{ab}(p)$ is a symmetric matrix, i.e. $\Gamma^{12}(p)=\Gamma^{21}(p)$, then 
the two equations (15) and (17) will combine into the single one 
\begin{equation}
\Gamma^{11}(p)+\Gamma^{22}(p)+2\cosh(\beta p^0/2)\Gamma^{12}(p)=0.
\end{equation}
Since both the thermal transformation equation of the four-point amputated function 
matrix $\Gamma_4^{aabb} \ (a,b=1,2)$ in a NJL model and the relations obeyed by its 
elements can be reduced to the ones of the two-point functions $\Gamma^{ab}(p)$, this 
makes it possible to discuss the thermal transformations of these four-point 
amputated functions by the same way to deal with two-point functions.
\section{Diagonalization in the $F\bar{F}$ basis of the matrix propagator for scalar 
bound state}
The propagator for the scalar bound state $(\bar{\psi}\psi)=\phi_S$ is a typical 
example of four-point amputated function in the NJL model.  Its effective $2\times 
2$ matrix form can be expressed by [12]
\begin{eqnarray}
\Gamma_S(p)&=&\left(\matrix{
\Gamma_S^{11}(p)& \Gamma_S^{12}(p) \cr
\Gamma_S^{21}(p)& \Gamma_S^{22}(p)\cr}\right)=
\frac{1}{|K(p)+H(p)-iS(p)|^2-R^2(p)} \nonumber \\
&&
\times\left(\matrix{\dfrac{S(p)-i[K^*(p)+H(p)]}{p^2-4m^2+i\varepsilon}
              &\dfrac{-(p^2-4m^2)R(p)}{( p^2-4m^2)^2+\varepsilon^2}\cr
               \dfrac{-(p^2-4m^2)R(p)}{( p^2-4m^2)^2+\varepsilon^2}
              &\dfrac{S(p)+i[K(p)+H(p)]}{p^2-4m^2-i\varepsilon}\cr}
         \right),
\end{eqnarray}
where $K(p)$ is the zero-temperature loop integral and can be divided into real and 
imaginary part, i.e. $K(p)={\rm Re}K(p)+i{\rm Im}K(p)$, $H(p)$, $S(p)$ and $R(p)$ 
are all real and even functions of $p$.  The explicit expressions of these functions 
will not given here because they are not important for discussions of thermal 
transformation except the relation [12]
\begin{equation}
S'(p)=S(p)-{\rm Im}K(p)=R(p)\cosh(\beta p^0/2).
\end{equation}
It is seen from Eq. (19) that the matrix elements $\Gamma^{ab}(p)$ obey the relations
\begin{equation}
\Gamma_S^{22}(p)=[\Gamma_S^{11}(p)]^*, \ \ 
\Gamma_S^{21}(p)= \Gamma_S^{12}(p)= {\Gamma_S^{12}(p)}^*,
\end{equation}
which include five relations among the eight real components of $\Gamma^{ab}(p)$, 
thus only three of $\Gamma^{ab}(p)$ are independent. Eq. (21) can be combined into
\begin{equation}
\Gamma_S^{11}(p)+\Gamma_S^{22}(p)+A \ \Gamma_S^{12}(p)=0.
\end{equation}
By means of Eqs. (19) and (20) we will obtain
\begin{eqnarray}
A&=&-2\frac{{\rm Re}\Gamma_S^{11}(p)}{\Gamma_S^{12}(p)}
                 =2\left[\frac{S'(p)}{R(p)}-\varepsilon\frac{{\rm Re}K(p)+H(p)}{(p^2-4m^2)R(p)}\right]\nonumber \\
&=&2\cosh(\beta p^0/2)-2\varepsilon\frac{{\rm Re}K(p)+H(p)}{(p^2-4m^2)R(p)}.
\end{eqnarray}
In Eq. (23) the term containing $\varepsilon =0_+$ only affects the position of the 
pole $p^2=4m^2$; if it is ignored, then Eq. (22) will become
\begin{equation}
\Gamma_S^{11}(p)+\Gamma_S^{22}(p)+2\cosh(\beta p^0/2)\Gamma_S^{12}(p)=0.
\end{equation}
This implies that $\Gamma_S^{ab}(p) \ (a,b=1,2)$ in Eq. (19) does satisfy the relation 
(18).  \\ 
\indent In view of Eq. (24), one may always find a transformation matrix $O(p)$ which 
will make some elements of the transformed matrix $\hat{\Gamma}_S(p)$ are identical 
to zeroes [9]. The diagonalization in the $F\bar{F}$ basis means that we can seek 
a $2\times 2$ matrix $O(p)$ which, through the equation like Eq. (8), make 
$\hat{\Gamma}_S(p)$
become a diagonal matrix, i.e. 
\begin{equation}
O(p)\Gamma_S(p)O^T(-p)=\hat{\Gamma}_S(p)= \left(\matrix{
\Gamma^{\phi_S}(p)& 0\cr
0& {\Gamma^{\phi_S}}^*(p)\cr}\right).
\end{equation}
Assume that 
\begin{equation}
O(p)=\left(\matrix{a(p) & b(p) \cr
                     c(p) & d(p) \cr}\right),
\end{equation}
then, noting that $\Gamma_S^{21}(p)= \Gamma_S^{12}(p)$ in Eq. (21), we will obtain 
respectively from \\
$[O(p)\Gamma_S(p)O^T(-p)]^{12}=0$ and
$[O(p)\Gamma_S(p)O^T(-p)]^{21}=0$ that 
\begin{equation}
a(p)c(-p)\left\{\Gamma_S^{11}(p)+\left[\frac{d(-p)}{c(-p)}
+\frac{b(p)}{a(p)}\right]\Gamma_S^{12}(p)
+\frac{b(p)}{a(p)}\cdot \frac{d(-p)}{c(-p)}\Gamma_S^{22}(p)\right\}=0
\end{equation}
and 
\begin{equation}
c(p)a(-p)\left\{\Gamma_S^{11}(p)+\left[\frac{b(-p)}{a(-p)}
+\frac{d(p)}{c(p)}\right]\Gamma_S^{12}(p)
+\frac{d(p)}{c(p)} \frac{b(-p)}{a(-p)}\Gamma_S^{22}(p)\right\}=0.
\end{equation}
Comparing Eq. (27) with Eq. (24), we will have 
\[
\frac{d(-p)}{c(-p)}=\alpha(p), \ \frac{b(p)}{a(p)}=\gamma(p)=2\cosh(\beta 
p^0/2)-\alpha(p), \ \frac{b(p)}{a(p)}\cdot \frac{d(-p)}{c(-p)}=\gamma(p)\cdot 
\alpha(p)=1,
\]
which result in that
\begin{equation}
\alpha(p) =e^{\beta|p^0|/2}, \ \ \gamma(p) =e^{-\beta|p^0|/2}.
\end{equation}
Comparing Eq. (28) with Eq. (24), we will have 
\[
\frac{d(p)}{c(p)}=\alpha(p), \ \frac{b(-p)}{a(-p)}=\gamma(p)=2\cosh(\beta 
p^0/2)-\alpha(p), \ \frac{b(-p)}{a(-p)}\cdot \frac{d(p)}{c(p)}=\gamma(p)\cdot 
\alpha(p)=1,
\]
which lead to Eq. (29) once again. The above results can be summarized as 
\begin{equation}
\frac{d(p)}{c(p)}=\frac{d(-p)}{c(-p)}=\alpha(p), \ 
\frac{b(p)}{a(p)}= \frac{b(-p)}{a(-p)}=\gamma(p),
\end{equation}
which are consistent with the fact that $\alpha(p)$ and $\gamma(p)$ are both even 
functions of $p$. Furthermore, owing to $\alpha(p)\cdot \gamma(p)=1$, we will have
\begin{equation}
b(p)d(p)=a(p)c(p).
\end{equation}
On the other hand, from Eq. (25) we must have 
\begin{equation}
[O(p)\Gamma_S(p)O^T(-p)]^{11}={[O(p)\Gamma_S(p)O^T(-p)]^{22}}^*,
\end{equation}
where
\begin{eqnarray}
[O(p)\Gamma_S(p)O^T(-p)]^{11}&=&a(p)a(-p)\left[
\Gamma_S^{11}(p)+\frac{b(-p)}{a(-p)}\Gamma_S^{12}(p)+ \frac{b(p)}{a(p)}\Gamma_S^{21}(p)\right.\nonumber\\
&&\left.+\frac{b(p)}{a(p)}\cdot\frac{b(-p)}{a(-p)}\Gamma_S^{22}(p) \right]\nonumber \\
&=& a(p)a(-p)\left[
\Gamma_S^{11}(p)+\gamma(p)\Gamma_S^{12}(p)+ \gamma(p)\Gamma_S^{21}(p)+\gamma^2(p)\Gamma_S^{22}(p) \right].\nonumber \\
\end{eqnarray}
It is incidentally indicated that when comparing Eq. (28) with Eq. (24),
it is also possible to take $b(-p)/a(-p)=\alpha(p)$ and $d(p)/c(p)=\gamma(p)$ ,
however, this will lead to $[O(p)\Gamma_S(p)O^T(-p)]^{11}= a(p)a(-p)\left[
\Gamma_S^{11}(p)+2\cosh(\beta p^0/2)\Gamma_S^{12}(p)+\Gamma_S^{22}(p) \right]=0$ 
due to Eq. (24), obviously inconsistent with the requirement of Eq. (25).  Next,
\begin{eqnarray*}
[O(p)\Gamma_S(p)O^T(-p)]^{22}&=&d(p)d(-p)\left[\frac{c(p)}{d(p)}\cdot \frac{c(-p)}{d(-p)}\Gamma_S^{11}(p)+\frac{c(p)}{d(p)}\Gamma_S^{12}(p)\right.\nonumber \\
&&\left.+ \frac{c(-p)}{d(-p)}\Gamma_S^{21}(p)+\Gamma_S^{22}(p)\right] \nonumber \\
&=&d(p)d(-p)\left[
\frac{1}{\alpha^2(p)}\Gamma_S^{11}(p)+\frac{1}{\alpha(p)}\Gamma_S^{12}(p)+ \frac{1}{\alpha(p)}\Gamma_S^{21}(p)+\Gamma_S^{22}(p)\right] \nonumber \\
&=&d(p)d(-p)\left[
\gamma^2(p)\Gamma_S^{11}(p)+\gamma(p)\Gamma_S^{12}(p)+ \gamma(p)\Gamma_S^{21}(p)+\Gamma_S^{22}(p)\right]
\end{eqnarray*}
Assume that $a(p)$, $b(p)$, $c(p)$ and $d(p)$ are all real functions of $p$, then 
by means of $\Gamma_S^{22}(p)={\Gamma_S^{11}}^*(p)$ in Eq. (21), Eq. (32)
will lead to
\begin{equation}
a(p)a(-p)=d(p)d(-p).
\end{equation}
From Eq. (30), it is reasonable to assume that $a(p)$, $b(p)$, $c(p)$ and $d(p)$ are 
all even functions of $p$, hence Eq. (34) is reduced to 
\[
a^2(p)=d^2(p).
\]
Taking $a(p)=d(p)$ and then from Eq. (31), we  will have $b(p)=c(p)$.  Consequently,
the transformation matrix $O(p)$ in Eq. (26) can be written by
\begin{equation}
O(p)=\left(\matrix{a(p) & b(p) \cr
                     b(p) & a(p) \cr}\right)
\end{equation}
If further assuming that the determinant of the matrix $O(p)$ is unity, i.e. 
${\rm det} O(p)=a^2(p)-b^2(p)=1$, then we can obtain from Eqs. (29) and (30) that
\begin{equation}
a^2(p)=1/[1-\gamma^2(p)]=n(p^0)+1, \ \ b^2(p)=n(p^0)=1/(e^{\beta|p^0|}-1).
\end{equation}
Then $O(p)$ will take the following form:
\begin{equation}
O(p)=\left(\matrix{\cosh\Theta & \sinh\Theta \cr
                    \sinh\Theta & \cosh\Theta \cr}\right), \ \sinh\Theta=[n(p^0)]^{1/2}=b, \ \cosh\Theta=[n(p^0)+1]^{1/2}=a.
\end{equation}
Therefore, the resulting transformation matrix $O(p)$ in the $F\bar{F}$ basis is just 
the thermal transformation matrix $M$ of the two-point function matrix for a real 
scalar field in the usual real-time formalism [6]. The form of $M$ was given in Ref. 
[12], but the details of its derivation were omitted there. From Eq. (33), the physical 
causal propagator for the scalar bound state $\phi_S$ will be
\begin{eqnarray}
\Gamma^{\phi_S}(p)&=&[O(p)\Gamma_S(p)O^T(-p)]^{11} \nonumber \\
&=&a^2(p)\Gamma_S^{11}(p)+2a(p)b(p)\Gamma_S^{12}(p)+
                       b^2(p)\Gamma_S^{22}(p) \nonumber \\
&=&[a^2(p)+b^2(p)]{\rm Re}\Gamma_S^{11}(p)+i{\rm Im}\Gamma_S^{11}(p)+ 
                      2a(p)b(p)\Gamma_S^{12}(p)\nonumber \\
&=&\frac{\cosh(\beta|p^0|/2)}{\sinh(\beta|p^0|/2)} {\rm Re}\Gamma_S^{11}(p)+i{\rm Im}\Gamma_S^{11}(p)+\frac{1}{\sinh(\beta|p^0|/2)}\Gamma_S^{12}(p),
\end{eqnarray}
where Eq. (37) has been used. By means of Eqs. (19) and (20), we can obtain from Eq. 
(38)
\begin{eqnarray}
\Gamma^{\phi_S}(p)&=&\frac{-i}{[{\rm Re}K(p)+H(p)]^2+R^2(p)\sinh^2(\beta|p^0|/2)}\cdot \frac{1}{(p^2-4m^2)^2+\varepsilon^2} \nonumber \\
&&\times\left(p^2-4m^2-i\varepsilon\frac{\cosh(\beta|p^0|/2)}{\sinh(\beta|p^0|/2)}\right)
[{\rm Re}K(p)+H(p)+iR(p)\sinh(\beta|p^0|/2)]\nonumber \\
&=&-i/[{\rm Re}K(p)+H(p)-iR(p)\sinh(\beta|p^0|/2)](p^2-4m^2+i\varepsilon)
\end{eqnarray}
which is precisely the physical causal propagator for the scalar bound state $\phi_S$
given in Ref. [12] and can be obtained from the Matsubara frequency $\Omega_m$'s 
analytic continuations $-i\Omega_m\to p^0+i\varepsilon p^0$ in the imaginary-time 
formalism.
\section{Transformation in the $RA$ basis of the matrix propagator for scalar bound 
state}
For fixing the positions of the propagators' poles more rigorously, we will use the 
relation (22) instead Eq. (24).  Owing to this relation among the matrix elements
$\Gamma_S^{ab}(p) \ (a, b=1,2)$, we may also seek a transformation matrix 
$\bar{O}(p)$ which makes all the diagonal elements of the transformed matrix 
$\hat{\bar{\Gamma}}_S(p)$ be equal to zeroes.  This is so called the transformation 
in the $RA$ basis [9].  Its explicit form is
\begin{equation}
\bar{O}(p)\Gamma_S(p)\bar{O}^T(-p)=\hat{\bar{\Gamma}}_S(p)=\left(\matrix{
                             0 & \Gamma_a^{\phi_S}(p) \cr
                    \Gamma_r^{\phi_S}(p) & 0  \cr}\right).
\end{equation}
Assume that
\begin{equation}
\bar{O}(p)=\left(\matrix{\bar{a}(p) & \bar{b}(p) \cr
                     \bar{c}(p) & \bar{d}(p) \cr}\right),
\end{equation}
then, owing to $\Gamma_S^{21}(p)= \Gamma_S^{12}(p)$, Eq. (40) indicates that
\begin{eqnarray}
[\bar{O}(p)\Gamma_S(p)\bar{O}^T(-p)]^{11}&=&\bar{a}(p)\bar{a}(-p)\left\{
\Gamma_S^{11}(p)+\left[\frac{\bar{b}(-p)}{\bar{a}(-p)}+ \frac{\bar{b}(p)}{\bar{a}(p)}\right]\Gamma_S^{12}(p)
+\frac{\bar{b}(p)}{\bar{a}(p)}\cdot\frac{\bar{b}(-p)}{\bar{a}(-p)}\Gamma_S^{22}(p) \right\}\nonumber \\
&=&0
\end{eqnarray}
and
\begin{eqnarray}
[\bar{O}(p)\Gamma_S(p)\bar{O}^T(-p)]^{22}&=&\bar{c}(p)\bar{c}(-p)\left\{\Gamma_S^{11}(p)+\left[\frac{\bar{d}(-p)}{\bar{c}(-p)}+ \frac{\bar{d}(p)}{\bar{c}(p)}\right]\Gamma_S^{12}(p)
+\frac{\bar{d}(p)}{\bar{c}(p)}\cdot\frac{\bar{d}(-p)}{\bar{c}(-p)}\Gamma_S^{22}(p)\right]\nonumber \\
&=&0.
\end{eqnarray}
For the use of Eq. (22), we may assume that
\begin{equation}
\frac{\bar{b}(\pm p)}{\bar{a}(\pm p)}=e^{\pm \eta(p^0)\bar{\Theta}}, \ \frac{\bar{d}(\pm p)}{\bar{c}(\pm p)}=e^{\mp \eta(p^0)\bar{\Theta}}, \  \eta(p^0)=\frac{p^0}{|p^0|},
\end{equation}
where
\begin{equation}
2\cosh\bar{\Theta}=A.
\end{equation}
It is indicated that $\bar{d}(\pm p)/\bar{c}(\pm p)=e^{\pm \eta(p^0)\bar{\Theta}}$, 
even though allowed by Eq. (43), should be removed because it will lead to 
$[\bar{O}(p)\Gamma_S(p)\bar{O}^T(-p)]^{12}=
[\bar{O}(p)\Gamma_S(p)\bar{O}^T(-p)]^{21}=0$ due to Eq. (22) and these do not 
satisfy the requirement of Eq. (40). Substituting Eq. (44) into Eq. (42), we can obtain 
the relations
\begin{equation}
\Gamma_S^{22}(p)+e^{-\eta(p^0)\bar{\Theta}}\Gamma_S^{12}(p)=-\left[\Gamma_S^{11}(p)+e^{\eta(p^0)\bar{\Theta}}\Gamma_S^{12}(p)\right]
\end{equation}
or
\begin{equation}
\Gamma_S^{22}(p)+e^{\eta(p^0)\bar{\Theta}}\Gamma_S^{12}(p)=-\left[\Gamma_S^{11}(p)+e^{-\eta(p^0)\bar{\Theta}}\Gamma_S^{12}(p)\right]
\end{equation}
By means of Eqs. (40),(41) and (44), we have
\begin{eqnarray}
\Gamma_a^{\phi_S}(p)&=& [\bar{O}(p)\Gamma_S(p)\bar{O}^T(-p)]^{12} \nonumber \\
                    &=&\bar{a}(p)\bar{c}(-p)\left[
\Gamma_S^{11}(p)+\frac{\bar{d}(-p)}{\bar{c}(-p)}\Gamma_S^{12}(p)+ \frac{\bar{b}(p)}{\bar{a}(p)}\Gamma_S^{21}(p)+\frac{\bar{b}(p)}{\bar{a}(p)}\cdot\frac{\bar{d}(-p)}{\bar{c}(-p)}\Gamma_S^{22}(p) \right]\nonumber \\
                    &=&\bar{a}(p)\bar{c}(-p)\left[
\Gamma_S^{11}(p)+2e^{\eta(p^0)\bar{\Theta}}\Gamma_S^{12}(p)+ e^{2\eta(p^0)\bar{\Theta}}\Gamma_S^{22}(p) \right]\nonumber \\
                    &=&\bar{a}(p)\bar{c}(-p)\left[1- e^{2\eta(p^0)\bar{\Theta}}\right]\left[
\Gamma_S^{11}(p)+e^{\eta(p^0)\bar{\Theta}}\Gamma_S^{12}(p)\right],
\end{eqnarray}
where Eq. (46) has been used and
\begin{eqnarray}
\Gamma_r^{\phi_S}(p)&=& [\bar{O}(p)\Gamma_S(p)\bar{O}^T(-p)]^{21} \nonumber \\
                    &=&\bar{c}(p)\bar{a}(-p)\left[
\Gamma_S^{11}(p)+\frac{\bar{b}(-p)}{\bar{a}(-p)}\Gamma_S^{12}(p)+ \frac{\bar{d}(p)}{\bar{c}(p)}\Gamma_S^{21}(p)+\frac{\bar{d}(p)}{\bar{c}(p)}\cdot\frac{\bar{b}(-p)}{\bar{a}(-p)}\Gamma_S^{22}(p) \right]\nonumber \\
                    &=&\bar{c}(p)\bar{a}(-p)\left[
\Gamma_S^{11}(p)+2e^{-\eta(p^0)\bar{\Theta}}\Gamma_S^{12}(p)+
 e^{-2\eta(p^0)\bar{\Theta}}\Gamma_S^{22}(p) \right]\nonumber \\
                    &=&-\bar{c}(p)\bar{a}(-p)
\left[1-e^{-2\eta(p^0)\bar{\Theta}}\right]\left[
\Gamma_S^{22}(p)+e^{\eta(p^0)\bar{\Theta}}\Gamma_S^{12}(p)\right],
\end{eqnarray}
where Eq. (47) has been used. Now define
\begin{equation}
\Gamma_a^{\phi_S}(p)\equiv \Gamma_S^{11}(p)+e^{\eta(p^0)\bar{\Theta}}\Gamma_S^{12}(p)
\end{equation}
and
\begin{equation}
\Gamma_r^{\phi_S}(p)\equiv
 -[\Gamma_S^{22}(p)+e^{\eta(p^0)\bar{\Theta}}\Gamma_S^{12}(p)]
\end{equation}
which obey the relation
\begin{equation}
[\Gamma_a^{\phi_S}(p)]^*=-\Gamma_r^{\phi_S}(p),
\end{equation}
then from Eqs. (48) and (49) we are led to that
\[\bar{a}(p)\bar{c}(-p)\left[1- e^{2\eta(p^0)\bar{\Theta}}\right]=1,\]
\begin{equation}
\bar{c}(p)\bar{a}(-p)
\left[1-e^{-2\eta(p^0)\bar{\Theta}}\right]=1.
\end{equation}
The six independent relations contained in Eqs. (44) and (53) allow us to determine 
the eight unknown $\bar{a}(\pm p)$, $\bar{b}(\pm p)$, $\bar{c}(\pm p)$ and $\bar{d}(\pm p)$ up to their being 
expressed by $\bar{a}(p)$ and $\bar{a}(-p)$. The results are as follows.
\[\bar{b}(\pm p)=e^{\pm \eta(p^0)\bar{\Theta}}\bar{a}(\pm p),\ 
\bar{c}(\pm p)= e^{\pm \eta(p^0)\bar{\Theta}}\bar{d}(\pm p), \]
\[
\bar{d}(\pm p)=\pm \frac{1}{2\eta(p^0)\sinh\bar{\Theta}\bar{a}(\mp p)}.
\]
As a result, the transformation matrix $\bar{O}(p)$ in the $RA$ basis may be written 
by 
\begin{equation}
\bar{O}(p)=\frac{1}{\bar{a}(-p)}\left(\matrix{
   \bar{a}(p)\bar{a}(-p) &e^{\eta(p^0)\bar{\Theta}}\bar{a}(p)\bar{a}(-p) \cr
e^{\eta(p^0)\bar{\Theta}}/2\eta(p^0)\sinh\bar{\Theta} &1/2\eta(p^0)\sinh\bar{\Theta} \cr}\right)
\end{equation}
which will give the transformed result in Eq. (40). From Eqs. (45) and (23), we have
\begin{equation}
e^{\eta(p^0)\bar{\Theta}}=\cosh(\beta p^0/2)-\varepsilon \frac{{\rm Re}K(p)+H(p)}{(p^2-4m^2)R(p)}+\eta(p^0)\sinh(\beta|p^0|/2),
\end{equation}
thus $\Gamma_a^{\phi_S}(p)$ and $\Gamma_r^{\phi_S}(p)$ have the following 
expressions:
\begin{eqnarray}
\Gamma_a^{\phi_S}(p)&\equiv &\Gamma_S^{11}(p)+e^{\eta(p^0)\bar{\Theta}}\Gamma_S^{12}(p) \nonumber \\
&=&\frac{1}{[{\rm Re}K(p)+H(p)]^2+R^2(p)\sinh^2(\beta|p^0|/2)}\cdot \frac{1}{(p^2-4m^2)^2+\varepsilon^2}\nonumber \\
&&\times \left(\frac{{}}{{}}\{S'(p)-i[{\rm Re}K(p)+H(p)]\}(p^2-4m^2-
i\varepsilon)\right.\nonumber \\
&&\left.-\left[
\cosh(\beta p^0/2)-\varepsilon \frac{{\rm Re}K(p)+H(p)}{(p^2-4m^2)R(p)}+\eta(p^0)\sinh(\beta|p^0|/2)\right](p^2-4m^2)R(p)\right) \nonumber \\
&=&\frac{1}{[{\rm Re}K(p)+H(p)]^2+R^2(p)\sinh^2(\beta|p^0|/2)}\cdot \frac{-i}{(p^2-4m^2)^2+\varepsilon^2}\nonumber \\
&&\times\{(p^2-4m^2)[{\rm Re}K(p)+H(p)-i\eta(p^0)R(p)\sinh(\beta|p^0|/2)]+\varepsilon R(p)\cosh(\beta p^0/2) \} \nonumber \\
&=& \frac{-i}{{\rm Re}K(p)+H(p)+i\eta(p^0)R(p)\sinh(\beta|p^0|/2)}\cdot \frac{1}{(p^2-4m^2)^2+\varepsilon^2}\nonumber \\
&&\times \left\{p^2-4m^2+i\varepsilon \eta(p^0)\frac{R^2(p)\sinh(\beta|p^0|/2) \cosh(\beta p^0/2)}
{[{\rm Re}K(p)+H(p)]^2+R^2(p)\sinh^2(\beta|p^0|/2)}\right\} \nonumber \\
&=&-i/[{\rm Re}K(p)+H(p)+iR(p)\sinh(\beta p^0/2)]
                           [p^2-4m^2-i\varepsilon \eta(p^0)],
\end{eqnarray}
and from Eq. (52),
\begin{equation}
\Gamma_r^{\phi_S}(p)= -i/[{\rm Re}K(p)+H(p)-iR(p)\sinh(\beta p^0/2)]
                           [p^2-4m^2+i\varepsilon\eta(p^0)].
\end{equation}
Therefore, through the transformation (40) in the $RA$ basis, we have just obtained 
the advanced and retarded propagator $\Gamma_a^{\phi_S}(p)$ and 
$\Gamma_r^{\phi_S}(p)$ for the scalar bound state $\phi_S$. It is indicated that the 
rigorous definition (55) of $e^{\eta(p^0)\bar{\Theta}}$ ensures the correct 
positions of the poles of $\Gamma_a^{\phi_S}(p)$ and $\Gamma_r^{\phi_S}(p)$. These 
results coincide with the ones obtained in the imaginary-time formalism respectively 
by the Matsubara frequency $\Omega_m$'s analytic continuations $-i\Omega_m\to 
p^0-i\varepsilon$ and $-i\Omega_m\to p^0+i\varepsilon$ [12].
\section{Conclusions}
Based a general analysis of four-point Green functions in the real-time thermal field 
theory, we have proven that the calculation of the four-point amputated functions 
in a NJL model in the fermion bubble diagram approximation is equivalent to the one 
of two-point functions in the following meanings that the thermal transformation of 
a four-point amputated function matrix can be reduced to the one of corresponding 
two-point function matrix and the same reduction is true to the relations satisfied 
by the elements of the two matrices. We further discuss in detail the thermal 
transformations of the matrix propagator for a scalar bound state in the $F\bar{F}$ 
basis and in the $RA$ basis and find out corresponding transformation matrices.  In 
the former case, we obtain physical causal propagator and in the latter case , physical 
advanced and retarded propagator. All the results coincide with the ones derived in 
the imaginary-time formalism by analytic continuations of the discrete Matsubara 
frequency.  This shows once again complete equivalence of the two formalisms of 
thermal field theory on the problem of the propagators for scalar bound states in the NJL model.

\end{document}